# Folding Thermodynamics of Model Four-strand Antiparallel $\beta$–sheet Proteins

Hyunbum Jang[*], Carol K Hall[*] and Yaoqi Zhou[+]

[*]*Department of Chemical Engineering, North Carolina State University, Raleigh, North Carolina 27695-7905;*
[+]*Department of Physiology and Biophysics, State University of New York at Buffalo, 124 Sherman Hall, Buffalo, New York 14214*

**ABSTRACT**  The thermodynamic properties for three different types of off-lattice four-strand $\beta$–sheet protein models interacting via a hybrid Go-type potential have been investigated. Discontinuous molecular dynamic simulations have been performed for different sizes of the bias gap $g$, an artificial measure of a model protein's preference for its native state. The thermodynamic transition temperatures are obtained by calculating the squared radius of gyration $R_g^2$, the root-mean-squared pair separation fluctuation $\Delta_B$, the specific heat $C_v$, the internal energy of the system $E$, and the Lindemann disorder parameter $\Delta_L$. In spite of the simplicity, the protein-like heteropolymers have shown a complex set of protein transitions as observed in experimental studies. Starting from high temperature, these transitions include a collapse transition, a disordered-to-ordered globule transition, a folding transition, and a liquid-to-solid transition. These transitions strongly depend on the native-state geometry of the model proteins and the size of the bias gap. A strong transition from the disordered globule state to the ordered globule state with large energy change and a weak transition from the ordered globule state to the native state with small energy change were observed for the large gap models. For the small gap models no native structures were observed at any temperature, all three $\beta$–sheet proteins fold into a partially-ordered globule state which is geometrically different from the native state. For small bias gaps at even lower temperatures, all protein motions are frozen indicating an inactive solid-like phase.

## INTRODUCTION

Proteins are macromolecules that fold into a specific native conformation to perform their biological function under physiological conditions. When they aggregate abnormally *in vivo*, however, the consequences are serious biomedical problems that include a host of ultimately fatal protein deposition diseases (Clark and Steele, 1992; Eaton and Hofrichter, 1990; Gallo et al., 1996; Massry and Glasscock, 1983; Moore and Melton, 1997). An example is Alzheimer's disease (Selkoe, 1991; Simmons et al., 1994); it is thought to be caused by the abnormal deposition of insoluble fibril plaques in the extracellular space of brain tissue during aging. The major component of plaques is a 39- to 43- residue $\beta$–peptide, called $\beta$–amyloid peptide (A$\beta$), whose predominant secondary structure in the fibril is a $\beta$–strand. Although recent extensive experimental studies (Benzinger et al., 1998, 2000; Burkoth et al., 1998; Esler et al., 1996, 2000; Lazo and Cowning, 1998; Lynn and Meredith, 2000; Sunde et al., 1997; Zhang et al., 2000) have gone a long way toward revealing the properties and physiological functions of $\beta$–amyloid peptide (A$\beta$), the causes of amyloid fibril formation are largely unknown and the fundamental mechanisms underlying aggregation of $\beta$–amyloid are still under investigation. Thus, it would be of interest to see if computer simulations can offer any insights into the basic principles behind folding and aggregation of peptides made up of $\beta$–strands.

The aim of this study is to investigate the thermodynamic properties of $\beta$–strand proteins. The ultimate goal of our work is to reveal the mechanisms underlying protein aggregation with particular focus on the role played by $\beta$–strands. In order to study protein aggregation by computer simulation, one must devise protein models that capture not only the basic physical features of real proteins, but also allow the simulation of many proteins with current computer capability. At present, protein models for computer simulations of folding can be divided mainly into two types: high-resolution models and low-resolution models. High-resolution or all-atom models (Brooks et al., 1983; Ferrara and Caflisch, 2000; Lazaridis and Karplus, 1997; Weiner et al., 1986) include detailed information on protein geometry and motion and yield useful information on the folding of a single real protein at short times. However computer limitations preclude simulating for more than one $\mu s$ (Duan and Kollman, 1998). Since aggregation time scales are considerably longer than folding time scales, all-atom representations are too cumbersome to allow for a meaningful description of the folding and aggregation of many chain systems.

In marked contrast to high-resolution models, low-resolution or simplified protein models offer the opportunity to study the behavior of multi-protein systems. These models provide numerous insights into the thermodynamic and kinetic properties of protein folding. The low-resolution models employed in simulation studies of protein folding can be divided into two classes: on-lattice (Chan and Dill, 1994; Dill and Stigter, 1995; Go and Taketomi, 1978, 1979; Gupta and Hall, 1997, 1998; Gupta et al., 1999; Kolinski et al., 1995; Lau and Dill, 1989; Miller et al., 1992; Skolnick and Kolinski, 1991; Taketomi et al., 1975; Ueda et al., 1978) and off-lattice models (Dokholyan et al., 1998, 2000; Guo and Thirumalai, 1995, 1996; Guo and Brooks III, 1997; Guo et al., 1997; Nymeyer et al., 1998; Pande and Rokhsar, 1998; Shea et al., 2000; Takada et al,. 1999; Zhou and Karplus, 1997a, 1997b, 1999). The first low-resolution lattice models of proteins were introduced by Go *et al.* (Go and Taketomi, 1978, 1979; Taketomi et al., 1975; Ueda et al., 1978) in the late 70s. In their models, the protein was treated as a sequence of beads on a two- or three-dimensional lattice with the native tertiary structure being determined by introducing a set of pre-assigned attractive residue-residue pairs at the nearest-neighbor lattice sites. Another class of lattice model used to describe protein folding is the HP lattice model proposed by Lau and Dill (1989). To mimic the hydrophobic effect, the protein is modeled as a chain of polar (P) and hydrophobic (H) residues with an attractive potential between non-bonded H beads and zero potential between non-bonded H and P or P and P residues. Simulation results for H-P models have provided useful information on protein folding-refolding pathways for isolated chains (Chan and Dill, 1994; Gupta and Hall, 1997; Gupta et al., 1999; Miller et al., 1992) and protein aggregation for multi-chain systems (Dill and Stigter, 1995; Gupta and Hall, 1998).

Recently, off-lattice models have gained attention as they can more accurately represent the real features of proteins. Guo *et al*. (Guo and Thirumalai, 1995, 1996; Guo and Brooks III, 1997; Guo et al., 1997) have investigated the thermodynamics and kinetics of protein folding using Langevin simulations on off-lattice protein models. Their heteropolymer models consist of $N$ connected beads which correspond to three types of residues: hydrophobic, hydrophilic, and neutral. They found two characteristic temperatures, a collapse transition temperature $T_\theta$ and a folding transition temperature $T_f$ in their thermodynamic studies of $\beta$–barrel and four-helix bundle proteins. Zhou and Karplus (1997a, 1997b, 1999) recently introduced an off-lattice heteropolymer model for a three-helix bundle protein with a hybrid Go-type potential. They used a bias gap parameter, $g$, to vary the strength of native contacts relative to that of non-native contacts. In spite of the model's simplicity, thermodynamic studies on this protein-like model exhibit complex protein transitions that have been observed experimentally including a collapse transition, a disordered globule to ordered globule transition, an ordered globule to native transition, and a transition to a surface frozen inactive state (Ptitsyn, 1995). In a study of the model's folding kinetics, Zhou and Karplus (1997a, 1997b, 1999) found that the larger the bias gap (closer it is to native state) the faster the protein folds. Well-defined folding pathways for the larger gap model showed a fast folding with no intermediates or slower folding with misfolded intermediates, while the smaller gap model exhibited long-lived intermediates with multiple pathways to the native state.

In this paper we investigate the thermodynamic phase behavior of three different four-strand antiparallel $\beta$–sheet peptides: the $\beta$–double-hairpin, the $\beta$–clip, and the $\beta$–twist. These $\beta$–strand peptides have different native state conformations, as shown in Fig. 1. This work is a first step in our effort to understand the aggregation of $\beta$–amyloid peptides. Our off-lattice heteropolymer models contain 39 connected beads, and each bead represents an amino acid residue. Non-bonded beads can interact through hybrid Go-type potential (Go and Taketomi, 1978, 1979; Taketomi et al., 1975; Ueda et al., 1978) modeled as a square-well or square-shoulder potential depending on the value of the bias gap parameter $g$. Discontinuous molecular dynamic (DMD) simulations (Alder and Wainwright, 1959; Rapaport, 1978; Smith et al., 1996) for bias gaps ranging from 0.3 to 1.5 were performed. The bias gap measures the difference in interaction strength between the native and non-native contacts: the larger it is the more the native state is favored over the non-native state. Intermediate values of the bias gap are thought to be the most representative of a real protein in the equilibrium and dynamic behavior. By exploring how variations in our model protein's bias gap, an artificial measure of its "preference" for the native state, influence the types of phase behavior observed, we can get a feeling for how a real protein's preference for its native state, as measured for example by the energy difference between the denatured and native state, is manifested in the protein's phase diagram and vice versa. Thermodynamic quantities such as the squared radius of gyration, $R_g^2$, the root-mean-squared pair separation fluctuation, $\Delta_B$, the specific heat, $C_v$, the internal energy, $E$, and the Lindemann disorder parameter, $\Delta_L$, were calculated during equilibrium simulations at several different temperatures.

Highlights of our results are the following. We find that the $\beta$–sheet proteins have complex protein transitions that are qualitatively similar to those seen in the thermodynamic studies of three-helix bundle proteins (Zhou and Karplus, 1997a, 1997b, 1999), even though the geometric structures of the two models are different. Starting from high temperature, these transitions include a collapse transition, a disordered-to-ordered globule transition, a folding transition, and a liquid-to-solid transition for all values of the bias gap. The high temperature transitions, i.e. the collapse transition and the disordered-to-ordered globule transition, exist for all three $\beta$–sheet proteins, although the native-state geometry of the three model proteins is different. However the low temperature transitions, i.e. the folding transition and the liquid-to-solid transition, strongly depend on the native-state geometry of the model proteins and the size of the bias gap. For small bias gaps at low temperatures, all three $\beta$–sheet proteins fold into a partially-ordered globule state which is geometrically different from the native state. For small bias gaps at even lower temperatures, all protein motions are frozen indicating an inactive solid-like phase. For large bias gaps at low temperatures, all three $\beta$–sheet proteins fold into the native state. For large bias gaps at even lower temperatures, the $\beta$–double-hairpin model exhibits no solid-like phase above $T^* = 0.07$, the lowest temperature studied, while the $\beta$–clip and $\beta$–twist models exhibit a solid-like phase for all values of $g$. The Lindemann criterion (Lindemann, 1910) suggests that the native state for the $\beta$–double-hairpin is a molten globule since all of the beads exhibit liquid-like motion. In contrast, the native states of the other two model proteins near the liquid-to-solid transition are surface-molten solids, since core beads exhibit inactive solid-like motion and surface beads exhibit liquid-like motion.

**THE MODELS AND SIMULATION METHOD**

We consider three different off-lattice protein models whose native states are four-strand antiparallel $\beta$–sheets. The global energy minimum structures for the three different $\beta$–sheet models are shown in Fig. 1. The "zig-zag" shaped native protein in Fig. 1(a) is called a $\beta$–double-hairpin; sometimes it is known as betabellin (Lim et al., 2000). The paper-clip and helix-shaped native structures in Fig. 1(b) and 1(c) are called $\beta$–clip and $\beta$–twist, respectively (Kolinski et al., 1995). A qualitative difference between the $\beta$–double-hairpin structure and the other two structures is immediately apparent. In the native state, the $\beta$–double-hairpin is two-dimensional (planar), while the $\beta$–clip and $\beta$–twist have well-defined three-dimensional four-barrel structures. The reduced squared radius of gyration per bead, $R_g^2 / \sigma^2 N$, the total number of native contacts, $N_{\text{native}}^{\text{total}}$, and the reduced squared end-to-end distance, $r_{1-39}^2 / \sigma^2$, for the three global energy minimum structures are summarized in Table 1. Here $N$ is the number of beads and $\sigma$ is the diameter of each bead along the chain. Note that chains with a large number of native contacts are generally more compact and consequently have a small value of the squared radius of gyration.

Each model protein contains 39 connected beads. Each bead represents an amino acid residue that can be regarded as being localized at the $C_\alpha$ atom. Non-bonded beads can interact with each other through a square-well or square-shoulder potential (Zhou et al., 1996,1997),

$$u_{ij}(r) = \begin{cases} \infty, & r < \sigma \\ B_{ij}\varepsilon, & \sigma < r < \lambda\sigma \\ 0, & r > \lambda\sigma \end{cases}, \tag{1}$$

where $\sigma$ is the bead diameter, $\varepsilon$ is an energy parameter with $\varepsilon > 0$, $\lambda\sigma$ is the square-well diameter with $\lambda = 1.5$ throughout, and $r$ is the distance between residues $i$ and $j$. The quantity $B_{ij}\varepsilon$, the square-well depth or square-shoulder height, represents the interaction strength between non-bonded residue pair $i$ and $j$ and is defined as

$$B_{ij}\varepsilon = \begin{cases} B_N\varepsilon, & \text{Native contact} \\ B_O\varepsilon, & \text{Non - native contact} \end{cases}, \tag{2}$$

where $B_N$ and $B_O$ are measures of the relative strengths of the energies associated with the native and non-native pair interactions in this Go-type potential (Go and Taketomi, 1978, 1979;

Taketomi et al., 1975; Ueda et al., 1978). Non-bonded pairs of beads that are in contact in the global energy minimum structure experience an attractive interaction, i.e. $B_N < 0$, when their square-wells overlap. On the other hand, non-bonded pairs of beads that are not in contact in the global energy minimum structure experience either an attractive interaction ($B_O < 0$) or a repulsive interaction ($B_O > 0$). The parameter $B_O$ can be either positive or negative depending on the size of the bias gap parameter $g$,

$$B_O = (1 - g)B_N, \qquad (B_N < 0, \ g > 0),  \qquad (3)$$

where $g$ is the bias gap (Zhou and Karplus, 1997a, 1997b, 1999). The bias gap measures the ratio of the interaction strength between the native contacts and non-native contacts. Note that for $g > 1$, $B_O > 0$ ; in this case non-native contacts are repulsive so that the native state structure is strongly favored over any non-native state structure. For $0 < g < 1$, $B_O < 0$ ; all non-bonded contact pairs are attractive but native contacts are always more favorable than non-native contacts. For $g = 0$, native and non-native contacts are equally favorable, $B_O = B_N$; the model reduces to a homopolymer. The bias gap is an artificial measure of a model protein's preference for its native state; in a real protein this preference for the native state might be measured for example by the energy difference between the native and non-native state.

The interaction between two bonded beads, $i$ and $i + 1$, is given by an infinitely deep square-well potential,

$$u_{i,i+1}^{\text{bond}}(r) = \begin{cases} \infty, & r < (1 - \delta)\sigma \\ 0, & (1 - \delta)\sigma < r < (1 + \delta)\sigma \\ \infty, & r > (1 + \delta)\sigma \end{cases} \qquad (4)$$

where $\delta$ is a flexibility parameter which controls the bond-length. The bond length between neighboring beads can be varied over a small distance between the infinitely high potential barriers at $(1-\delta)\sigma$ and $(1+\delta)\sigma$. This method for creating a flexible bond length was introduced by Bellemans *et al.* (1980), and is known as Bellemans bond. The flexible bond-length parameter is set to $\delta = 0.1$ through the paper.

Simulations were performed using the discontinuous molecular dynamics (DMD) algorithm (Alder and Wainwright, 1959; Rapaport, 1978; Smith et al., 1996). All simulations were started from randomly generated configurations which were obtained from self-avoiding random walks. The initial velocities were chosen at random from a Maxwell-Boltzmann distribution at the simulation temperature. This DMD algorithm is very efficient for calculating the properties of systems containing hard spheres or square-well spheres (Smith et al., 1996; Zhou et al., 1997) and is significantly faster than standard molecular dynamics on Lennard-Jones chains. The algorithm proceeds by searching for the next collision time and collision pair, advancing all beads in time to the next collision event, and then calculating the velocity change of the colliding pair. The DMD simulation was conducted in the canonical ensemble with a constant number of particles, volume, and temperature. In order to maintain constant temperature, the Andersen stochastic collision method (Andersen, 1980) was used. In this method the system's particles collide with imaginary or ghost particles which serve as an effective heat bath. In the collision with the imaginary particles, the velocity of each bead is re-assigned from a Maxwell-Boltzmann distribution at the simulation temperature. A reduced ghost particle density of $n_g{}^* \equiv n_g\sigma^3 = 0.1$, where $n_g$ is the number density of ghost particles, was used to ensure that 1% ~ 10% of the collisions were ghost particle collisions (Zhou and Karplus, 1999; Zhou et al., 1997).

In order to determine the location of the collapse transition, the mean-squared radius of gyration, $R_g{}^2$, was determined where

$$R_g^2 \equiv \left\langle \frac{1}{N}\sum_{i=1}^{N}\left[(x_i-x_c)^2 + (y_i-y_c)^2 + (z_i-z_c)^2\right]\right\rangle_{\text{cf}} , \qquad (5)$$

with $N$ equal to the number of beads, $x_i, y_i, z_i$ equal to the coordinates of bead $i$, $x_c, y_c, z_c$ equal to the center of mass coordinates of the chain, and $\langle\ \rangle_{\text{cf}}$ denoting a configurational average. To save computing time the summation is taken over the selected configurations that were sampled every 1000 collisions after equilibration. Other transitions were determined by calculating the reduced specific heat,

$$C_v^* \equiv \frac{C_v}{k_B} = \frac{\left\langle E^2\right\rangle - \left\langle E\right\rangle^2}{k_B^2 T^2} \qquad (6)$$

and the reduced internal energy,

$$E^* \equiv \frac{\left\langle E\right\rangle}{\varepsilon} \qquad (7)$$

where $k_B$ is Boltzmann's constant and $\langle\ \rangle$ denotes an average over all collisions after equilibration. In calculating the specific heat and energy according to Eqs. (6) and (7), the weighted histogram method (Ferrenberg and Swendsen, 1989; Zhou et al., 1997) was used. In this method, temperature-independent degeneracy factors are extracted from the energy probability distribution in simulation runs at different temperatures. Once the degeneracy factors are known, a partition function is obtained for any temperature, and consequently all thermodynamic quantities can be calculated. Details of the method were reported elsewhere (Zhou et al., 1997). By plotting $C_v^*$ versus $T^*$, where $T^*$ is the reduced temperature, $T^* \equiv k_B T/\varepsilon$, one can determine the equilibrium transitions from the locations of the peaks and plateaus.

The degree of bead mobility can be characterized by the root-mean-squared (rms) pair separation fluctuation, $\Delta_B$, defined by

$$\Delta_B \equiv \frac{2}{N(N-1)}\sum_{i<j}\left(\frac{\left\langle r_{ij}^2\right\rangle_{\text{cf}}}{\left\langle r_{ij}\right\rangle_{\text{cf}}^2} - 1\right)^{\frac{1}{2}}, \qquad (8)$$

where $r_{ij}$ denotes the separation between beads $i$ and $j$, and $\langle\ \rangle_{\text{cf}}$ denotes the configurational average. To characterize the liquid-to-solid transition of the system, the root-mean-squared fluctuation for bead $i$, $\Delta_{Li}$, defined to be

$$\Delta_{Li} \equiv \left( \frac{\left\langle \left( |\mathbf{r}_i| - \langle |\mathbf{r}_i| \rangle_{\text{cf}} \right)^2 \right\rangle_{\text{cf}}}{\sigma^2} \right)^{\frac{1}{2}}, \qquad (9)$$

yields the Lindemann disorder parameter,

$$\Delta_L = \left( \sum_i \frac{\Delta_{Li}^2}{N} \right)^{\frac{1}{2}} \qquad (10)$$

where bead positions, $\mathbf{r}_i$ ($i = 1$ to $N$), are obtained by shifting the center of mass coordinates to the origin. The quantity $\Delta_{Li}$ in Eq. (9) defines the Lindemann disorder parameter for an individual bead, and $\Delta_L$ in Eq. (10) defines the Lindemann disorder parameter. The Lindemann criterion (Lindemann, 1910) is applied to the system to determine the solid or liquid phase. A system is regarded as a solid if its Lindemann disorder parameter, $\Delta_L$, is in the range of 0.1 to 0.15 (Bilgram, 1987; Löwen, 1994; Stillinger, 1995; Zhou et al., 1999), while a substance with $\Delta_L > 0.15$ is considered liquid.

The progression of a conformation towards the native state can be monitored by introducing the fraction of native contacts formed (Lazaridis and Karplus, 1997; Šali et al., 1994), $Q$, defined by

$$Q = \frac{N_{\text{native}}}{N_{\text{native}}^{\text{total}}}, \qquad (11)$$

where $N_{\text{native}}$ represents the number of native contacts in a given conformation and $N_{\text{native}}^{\text{total}}$ is the total number of native contacts in the global energy minimum structure. The fraction of native contacts $Q$ has a value between 0 to 1. When $Q = 1$ the model system can be regarded as being in the native structure, while when $Q \rightarrow 0$ the chain becomes a random coil that indicates the denatured state.

Simulations were performed for up to $10^9$ collisions to ensure equilibration. Each simulation was started from a different initial configuration at the temperature of interest. Equilibrium averages were typically taken by discarding the first half of the simulation. All equilibrium results were averaged over at least four independent runs. For simulations at low temperature, a random initial configuration was generated at a relatively high temperature of $T^* = 1.0$, and then gradually annealed toward the target temperature after which the equilibrium simulation was performed. This procedure can prevent the system from becoming trapped in a metastable state, i.e. a local free-energy minimum. Simulations were performed for bias gaps in the range of $g = 0.3$ to 1.5 at reduced temperatures in the range $T^* = 0.07$ to 5.0.

## RESULTS AND DISCUSSION

### Thermodynamic properties of $\beta$–double-hairpin

DMD simulations were performed for the $\beta$–double-hairpin protein to investigate phase behavior as a function of the size of the bias gap. At bias gaps of $g = 0.3$, $0.7$, $0.9$, and $1.3$, thermodynamic averages including the squared radius of gyration $R_g^2$, the root-mean-squared pair separation fluctuation $\Delta_B$, the specific heat $C_v$, and the internal energy $E$ were calculated as a function of temperature. These results for $g = 0.3$, $0.7$, $0.9$, and $1.3$ are shown in Figs. 2(a), 2(b), 2(c), and 2(d), respectively. All quantities are described in terms of reduced units in the figures. Statistical averages were obtained over at least four independent simulations. The error bars are the standard deviation in the measured values and are only shown for values larger than the size of the symbol. The results for the specific heat and the energy were obtained using the same weighted histogram method (Ferrenberg and Swendsen, 1989) that was introduced in the study of homopolymers (Zhou et al., 1997). The equilibrium transition temperatures of the model can be identified from the maximum value of the temperature derivative of the squared radius of gyration, $dR_g^2/dT^*$, and/or from peaks or plateaus in the specific heat.

To determine the collapse transition, the temperature derivative of the squared radius of gyration, $dR_g^2/dT^*$, is calculated along a spline fit of the data in $R_g^2$ versus $T^*$ graphs as shown in Fig. 2. For a small bias gap, $g = 0.3$, in Fig. 2(a) the maximum value of $dR_g^2/dT^*$ occurs at $T^* = 1.47$ indicating the collapse transition temperature. For $g = 0.7$, $0.9$, and $1.3$ in Figs. 2(b), 2(c), and 2(d), they occur at $T^* = 1.0$, $0.95$, and $0.88$, respectively. The chain has a conformational change from a random coil to the disordered globule state at the collapse transition. The disordered globule has more native contacts than that for the random coil, but due to the disordered nature of the state, none of the structures appear to be a well ordered native-like. Additional evidence for the existence of the collapse transition at $g = 0.3$ is that the $\langle \Delta_B \rangle$ versus $T^*$ curve possesses a broad, but clear, maximum located at a temperature just above $T^* = 1.47$. In addition, a plateau in the specific heat is observed near this temperature. In contrast to the broad peak for $g = 0.3$ in the $\langle \Delta_B \rangle$ curve, for $g = 0.7$, $0.9$, and $1.3$, there is a well-defined peak in the $\langle \Delta_B \rangle$ curve that evidently supports the collapse transition. The collapse transition peak in the $\langle \Delta_B \rangle$ versus $T^*$ curve can be explained in the following way. The root-mean-squared pair separation fluctuation $\langle \Delta_B \rangle$ increases, as expected, as $T^*$ increases but then decreases above the collapse transition temperature because the stretched chain in the random coil state is constrained by chain connectivity, resulting in less fluctuations in the bead-bead separations

The transition from the disordered globule state to the ordered globule state is associated with a distinct peak in the specific heat. This transition is strong with large energy change indicating a large conformational change. For $g = 0.3$, the peak is found at a temperature of $T^* = 0.58$, far below the collapse transition. This indicates that the disordered globule is a stable state over a wide temperature range; a trend characteristic of smaller values of $g$. However, for $g = 1.3$ the peak is found at a temperature of $T^* = 0.80$, only slightly below the collapse transition temperature. The collapse transition will eventually coincide with the disordered-to-ordered transition for $g > 1.5$.

The folding transition from the ordered globule state to the native state can be identified from the plateaus in the specific heat. At this transition, there are no major changes in structure, but the chain has a subtle conformational change toward the native state. For the smaller gap models, $g = 0.3$, no evidence for the folding transition was observed. This is because for $g < 0.7$, a well-ordered native state does not appear to exist; instead kinked or rolled structures were observed. Consequently, no plateaus associated with the folding transition were observed for the smaller gap models. For the $g = 0.7$ model, although a peak in the specific heat is found at a temperature of $T^* = 0.35$, the peak is not related to the folding transition. It is associated with the transition from the ordered globule state to a more-collapsed ordered globule state that we will call the partially-ordered globule state. For the larger gap models, $g = 0.9$ and 1.3, the plateaus at $T^* = 0.40$ in the specific heat are associated with the folding transition. This folding transition temperature is much lower than the disordered-to-ordered globule transition. This indicates that the ordered globule is a stable state over a wide temperature range for the larger gap models.

The results for the $\beta$–double-hairpin that we have just described are summarized in Fig. 3 which shows the phases that occur in the space spanned by the reduced temperature $T^*$ and the bias gap $g$. The diagram shows a complex set of protein transitions that is qualitatively similar to those observed for a model three-helix bundle protein (Zhou and Karplus, 1997a, 1997b, 1999). However, some quantitative differences between the $\beta$–double-hairpin results and the three-helix bundle results are immediately apparent. This is indicated in the figure by drawing solid lines for transitions that are qualitatively the same transitions as those observed for the three-helix bundle protein and dotted lines for transitions which are not observed in the three-helix bundle protein. The upper line in the figure indicates the collapse transition; it separates the random coil phase from the disordered globule phase. Moving down in the figure, the second line from the top indicates the transition from the disordered globule to the ordered globule; it is obtained from the first peak in the specific heat. The third line from the top is a folding transition for $g > 0.7$, that separates the ordered globule from the native state and a collapse transition to a partially-ordered globule (different from the native state) for $0.3 < g < 0.7$. As described later in this paper the partially-ordered globule state contains kinked or rolled structures which are distinctly different from their native state conformation. Interestingly, for $0.7 < g < 0.9$, we observed a transition from the native state to the partially-ordered globule state that is not observed in the three-helix bundle protein. This transition is associated with the specific heat peak for $g = 0.7$ at $T^* = 0.35$ and the small specific heat peak for $g = 0.9$ at $T^* = 0.115$. For $g < 0.9$ a transition at low temperature to a solid phase; this is obtained through application of the Lindemann criterion as will be discussed later. No solid phase was observed for $g > 0.9$ at any temperature investigated. Two triple points are found: one at $g = 0.3$ and $T^* = 0.58$ where the disordered, ordered and partially-ordered globule phases meet, and one at $g = 0.7$ and $T^* = 0.35$ where the native, ordered and partially-ordered globule phases meet.

Folding of the $\beta$–double-hairpin strongly depends on the bias gap and temperature. A set of sample structures for the $\beta$–double-hairpin characteristic of the final states observed during the simulations is presented in Fig. 4 (a) for $g = 0.7$ at $T^* = 0.3$, 4(b) for $g = 0.7$ at $T^* = 0.6$, 4(c) for $g = 1.3$ at $T^* = 0.5$, and 4(d) for $g = 1.3$ at $T^* = 0.2$. The structure in Fig 4(a) corresponds to the partially-ordered globule, the kinked or rolled structure that appears at low temperature for $g < 0.7$. (Recall that no native state structures appear at any temperatures for $g < 0.7$.) The partially-ordered globule state occurs for smaller gap models because the attraction between non-native contacts disturbs the tendency to order into the native state at low temperatures. Examples of the ordered globules are shown in Figs. 4(b) and 4(c). The ordered globule has more native contacts than the disordered globule. As $g$ increases, the structure corresponding to the ordered globule becomes more native-like (compare Figs. 4(b) and 4(c)). The native state structure of the $\beta$–double-hairpin is shown in Fig. 4(d). Note that for the off-lattice $\beta$–double-hairpin models with $g \leq 0.7$ none of the structures are the global energy minimum state, but for $0.7 < g < 1.0$ the geometry in the native state is very similar to the global energy minimum structure shown in Fig. 1(a) so that the number of native contacts is very close to that of the global energy minimum structure. For highly optimized models, $g \geq 1.0$, the native state is the global energy minimum state (Zhou and Karplus, 1999).

It is interesting to consider the variation in structural properties as folding progress. Fig. 5 shows the average value of the fraction of global energy minimum contacts formed (Lazaridis and Karplus, 1997; Šali et al., 1994), $\langle Q \rangle$, as a function of temperature $T^*$ and bias gap $g$. The collapse transition occurs at $\langle Q \rangle \sim 0.2$ for $g = 0.3$ and at $\langle Q \rangle \sim 0.3$ for $g = 1.3$. The transition from the disordered globule to the ordered globule occurs at $\langle Q \rangle$ values in the range of $\langle Q \rangle = 0.4 \sim 0.5$ for all values of $g$. In the native state, we find that $\langle Q \rangle > 0.9$ for $g > 0.7$, while for the partially-ordered globule, $\langle Q \rangle$ has values in the range from 0.6 to 0.9 depending on the size of $g$. In the three-dimensional graph, the flat surface at the top of the rear edge corresponds to the native state with $\langle Q \rangle \sim 1$ for $g > 0.7$.

Further information on the nature of the transition at low temperature can be obtained by calculating the Lindemann disorder parameter (Lindemann, 1910). This parameter is often used to analyze the liquid-to-solid transition (Bilgram, 1987; Löwen, 1994; Stillinger, 1995; Zhou et al., 1999). Fig. 6(a) shows the average value of the Lindemann disorder parameter, $\langle \Delta_L \rangle$, as a function of temperature for two selected values of the bias gap, $g = 0.3$ and 1.3. Each result for $\Delta_L$ at given temperature is a sum of the average of the root-mean-squared fluctuations over all beads. A form of the Lindemann criterion for melting is adopted here in which systems with $\Delta_L < 0.15$ are considered solid-like, while systems with $\Delta_L > 0.15$ are considered liquid. The dotted line in Fig. 6(a) indicates the criterion for the liquid-to-solid transition. It can be seen that for $g = 1.3$, as was seen in the phase diagram, no solid phase is found at any investigated temperatures because $\Delta_L > 0.15$, whereas, for $g = 0.3$ the liquid-to-solid transition occurs at $T^* = 0.19$. This agrees well with the temperature at which the inflection in the root-mean-squared pair separation, $\langle \Delta_B \rangle$, curve is found in Fig 2(a). Figs. 6(b) and 6(c) show the Lindemann disorder parameter for the individual beads $i$, $\Delta_{Li}$, as a function of the reduced distance of bead $i$ from the origin, $|\mathbf{r}_i|/\sigma$, for 6(b) $g = 0.3$ at $T^* = 0.1$, 0.3, and 0.6, and 6(c) for $g = 1.3$ at $T^* = 0.2$, 0.6, and 0.8. Note that the bead positions, $\mathbf{r}_i$, are defined with respect to the origin at the chain's center of mass. In Fig. 6(b), which corresponds to $g = 0.3$, the three selected temperatures correspond to the three different phases of the chain: the solid-like phase, partially-ordered globule, and disordered globule, respectively. At high temperature, $T^* = 0.6$, a large degree of freedom in the bead motion is observed. At low temperature, $T^* = 0.1$, the small bead fluctuations and small bead-bead separations give rise to a solid-like phase for the chain. In Fig. 6(c), which corresponds to $g = 1.3$, the three selected temperatures correspond to the three different phases of the chain: the native state, ordered globule, and disordered globule, respectively. At high temperature, $T^* = 0.8$, the large bead fluctuations in the $\Delta_{Li}$ graph indicate that the chain undergoes a transition from the disordered globule to the ordered globule. At low temperature, $T^* = 0.2$, the results of the individual bead fluctuations correspond to the native state. The native protein is regarded as a surface-molten solid, which means that the interior of native protein is solid-like, but the surface is liquid-like phase (Zhou et al., 1999). However we find that the native $\beta$–double-hairpin at $T^* = 0.2$ is not a surface-molten solid, since all the values of $\Delta_{Li}$ are greater than 0.15, which means that the system is considered liquid. Therefore, the native $\beta$–double-hairpin is more like a molten globule since all of the beads exhibit liquid-like motion. The reason that the native $\beta$–double-hairpin is not a surface-molten solid is that it is planar, and consequently the interior beads have considerable freedom in their motion; they can move perpendicular to the molecular plane even in the native state. This can be regarded as a case in which the topology determines the phase behavior.

To characterize the nature of the transitions displayed in Fig. 3, the energy distribution in the transition region has been investigated. Fig. 7 shows the energy distribution 7(a) for $g = 0.9$ at $T^* = 0.7949$ (the disordered-to-ordered globule transition), 7(b) for $g = 0.9$ at $T^* = 0.3998$ (the folding transition), and 7(c) for $g = 0.3$ at $T^* = 0.5848$ (the disordered-to-partially-ordered globule transition). The bimodal distribution in Fig. 7(a) for $g = 0.9$ indicates that the transition from the disordered globule to the ordered globule is a two-state transition; the disordered and ordered globule states coexist at the transition temperature. This two-state transition can be regarded as a first-order-like transition, since as pointed out by Zhou and Karplus (1999), the logarithm of the energy distribution via the weighted histogram method is directly proportional to the free energy, and the existence of a bimodal shape in the free energy distribution verifies the first-order-like transition (Zhou et al., 1996,1997). Note that this two-state transition is associated with the distinct peak in the specific heat as seen in Figs. 2(c) and 2(d) which indicate that the transition from the disordered globule to the ordered globule is strong with large energy change. The transition into the ordered globule is a two-state only for $g \geq 0.9$. The folding transition in Fig. 7(b) for $g = 0.9$ is not two-state since no bimodal energy distribution exhibits. This indicates that the ordered globule and native states of the $\beta$–double-hairpin are geometrically similar to each other. Hence, the folding transition is weak with small energy change and is continuous between the two states. Note that this transition is associated with the plateaus in the specific heat as seen in Figs. 2(c) and 2(d). For the transition from the disordered globule state to the partially-ordered globule state in Fig. 7(c) at $g = 0.3$, the transition is continuous even though it is associated with a distinct peak in the specific heat as seen in Fig. 2(a). The lack of a bimodal energy distribution for $g = 0.3$ suggests that the disordered and partially-ordered globule structures are similar to each other.

**The models of $\beta$–clip and $\beta$–twist**

The thermodynamic averages, $\langle R_g^2/\sigma^2 N \rangle$, $\langle \Delta_B \rangle$, $\langle C_v^*/N \rangle$, and $\langle E^*/N \rangle$ for the $\beta$–clip and $\beta$–twist models are shown as a function of $T^*$ in Figs. 8(a) and 8(b) ($\beta$–clip), and 8(c) and 8(d) ($\beta$–twist). For brevity the figures only show results for two selected values of the bias gap, $g = 0.3$ and 1.3. For the small bias gap, $g = 0.3$, the results in Figs. 8(a) for the $\beta$–clip and 8(c) for the $\beta$–twist are qualitatively similar to those in Fig. 2(a) for the $\beta$–double-hairpin model at $g = 0.3$. However for the large bias gap, $g = 1.3$, the thermodynamic averages for both the $\beta$–clip model and the $\beta$–twist model (Figs. 8(b) and 8(d) respectively) are qualitatively different from those of the $\beta$–double-hairpin model at $g = 1.3$ in Fig. 2(d). The differences are that kinks in the $\langle R_g^2/\sigma^2 N \rangle$ and $\langle \Delta_B \rangle$ curves at low temperatures and a second peak in the specific heat in Figs. 8(b) and 8(d) are observed for both the $\beta$–clip model and the $\beta$–twist model, but not for the $\beta$–double-hairpin model in Fig. 2(d). The small peak in the specific heat is related to the folding transition, while the kinks in the $\langle R_g^2/\sigma^2 N \rangle$ and $\langle \Delta_B \rangle$ curves at low temperatures are connected with the liquid-to-solid transition. The reason for the increases in $\langle R_g^2/\sigma^2 N \rangle$ at low temperatures for both the $\beta$–clip model and the $\beta$–twist model at $g = 1.3$ is that the $R_g^2$ value of the native protein is larger than that of the ordered globule, indicating that the collapsed chain undergoes a subtle conformational change from the ordered globule to the native state by elongating its strand.

The Lindemann disorder parameters are shown as a function of temperature and the bias gap for the $\beta$–clip and $\beta$–twist in Fig. 9(a) and 9(b) respectively. It can be seen that for both models the locations of the kinks in the $\langle \Delta_L \rangle$ curves at low temperatures and large $g$ are consistent with those at low temperatures found in the $\langle R_g^2/\sigma^2 N \rangle$ and $\langle \Delta_B \rangle$ curves as seen in Figs. 8(b) and 8(d). The solid-like phase with $\Delta_L < 0.15$ is prevalent at low temperature for the $\beta$–clip and $\beta$–twist models. Sharp increases at high temperatures in the $\langle \Delta_L \rangle$ curves for the large gap models are also consistent with those found in the $\langle R_g^2/\sigma^2 N \rangle$ curves and with the peaks in the $\langle \Delta_B \rangle$ curves in Figs. 8(b) and 8(d) indicating the collapse transition.

Figures 10(a) and 10(b) show the Lindemann disorder parameter for an individual bead, $\Delta_{Li}$, 10(a) for the $\beta$–clip model with $g = 0.9$ at $T^* = 0.1$, 0.26, and 0.5, and 10(b) for the $\beta$–twist model with $g = 0.7$ at $T^* = 0.1$, 0.28, and 0.6. For both the $\beta$–clip and $\beta$–twist models, those three selected temperatures correspond to the three different phases of the chain: the solid-like phase, the native state, and the ordered globule, respectively. In the native states near the liquid-to-solid transition ($\beta$–clip with $g = 0.9$ at $T^* = 0.26$ in Fig. 10(a) and $\beta$–twist with $g = 0.7$ at $T^* = 0.28$ in Fig. 10(b)), we find that the interiors of the chains are solid-like with $\Delta_{Li} < 0.15$, but the surfaces are liquid-like with $\Delta_{Li} > 0.15$. Thus the native proteins for the $\beta$–clip and $\beta$–twist are surface-molten solids. At low temperature, $T^* = 0.1$, for both the $\beta$–clip and $\beta$–twist models, the bead motions are frozen with all $\Delta_{Li} < 0.1$ indicating an inactive solid-like phase.

Figures 11(a) for the $\beta$–clip and 11(b) for the $\beta$–twist summarize the protein phases that occur in the space spanned by the reduced temperature $T^*$ and the bias gap $g$. The qualitative differences between the phase behavior results for the $\beta$–clip and $\beta$–twist models and those for the $\beta$–double-hairpin model (Fig. 3) are immediately apparent. For examples, the native states for the $\beta$–clip and $\beta$–twist models exist at low temperatures for $g > 0.5$, while for the $\beta$–double-hairpin it exists at low temperatures for $g > 0.7$. For the $\beta$–clip and $\beta$–twist models at even low temperatures, the liquid-to-solid transition exists for all values of $g$, but for the $\beta$–double-hairpin at even low temperatures it exists only for $g < 0.9$. However, for all three $\beta$–sheet protein models, although not shown in the figures, a similarity has been found that for $g \geq 0.9$ the energy distribution for the disordered-to-ordered globule transition is bimodal, confirming that the transition is two-state, but the energy distribution for the folding transition is not bimodal, indicating that the folding transition is continuous.

## CONCLUSIONS

We have studied the folding thermodynamics for three different types of off-lattice four-strand $\beta$–sheet protein models: the $\beta$–double-hairpin, the $\beta$–clip, and the $\beta$–twist. Discontinuous molecular dynamic simulations of these models have been performed for different sizes of the bias gap $g$, a measure of the strength of the native contacts relative to that of the non-native contacts. Although all three $\beta$–sheet protein models are simple representations for a real protein, all model proteins have shown a complex set of protein transitions as observed in experimental studies (Ptitsyn, 1995). These transitions include the collapse transition, the disordered-to-ordered globule transition, the folding transition, and the liquid-to-solid transition for all values of the bias gap. A qualitative difference between the present models results and the three-helix bundle results (Zhou and Karplus, 1997a, 1997b, 1999) is the existence of a partially-ordered globule state, which was not observed in the three-helix bundle protein. This partially-ordered globule state exists for all three $\beta$–sheet proteins at small value of the bias gap since the attraction between non-native contacts disturbs the tendency to order into the native state at low temperatures.

The bias gap $g$ in the models is clearly seen to be an important factor in controlling the phase behavior of the model proteins. A strong transition from the disordered globule state to the ordered globule state with large energy change and a weak transition from the ordered globule state to the native state with small energy change were observed for the larger gap models, $g \geq 0.9$, while for the smaller gap models, $g < 0.7$ for the $\beta$–double-hairpin and $g < 0.5$ for the $\beta$–clip and $\beta$–twist, no native structures were observed at any temperature. The Lindemann criterion suggests that the native $\beta$–double-hairpin is a liquid-like molten globule, while the native states of the other two model proteins near the liquid-to-solid transition are surface-molten solids, i.e. having an inactive solid-like phase inside and liquid-like phase at the surface. The $\beta$–double-hairpin's lack of a three-dimensional core is the major fact that the native $\beta$–double-hairpin is not a surface-molten solid since the interior beads in its planar geometry have considerable freedom in their motion causing the liquid-like molten globule and no liquid-to-solid transition. Further studies on the folding kinetics of the present models and the aggregation of multi-chain systems are being conducted in order to further our understanding of the complex behaviors of protein systems.

This work was supported by the National Institutes of Health under grant number GM-56766 and the National Science Foundation under grant number CTS-9704044. The work at Buffalo is supported in part by a grant from HHMI to SUNY Buffalo.

**TABLE 1**

|  | $\beta$–double-hairpin | $\beta$–clip | $\beta$–twist |
|---|---|---|---|
| $R_g^2/\sigma^2 N$ | 0.2350 | 0.2164 | 0.2163 |
| $N_{\text{native}}^{\text{total}}$ | 75 | 125 | 130 |
| $r_{1-39}^2/\sigma^2$ | 9 | 1 | 3 |

**TABLE 1** The reduced squared radius of gyration per bead, $R_g^2/\sigma^2 N$, the total number of native contacts, $N_{\text{native}}^{\text{total}}$, and the reduced squared distance between beads number 1 and 39, $r_{1-39}^2/\sigma^2$, in the global energy minimum structure for the three different $\beta$–sheet models.

**FIGURE CAPTIONS**

**FIGURE 1** The global energy minimum structures for: (a) the $\beta$–double-hairpin, (b) the $\beta$–clip, and (c) the $\beta$–twist model proteins.

**FIGURE 2** The average values of the reduced squared radius of gyration per bead, $\langle R_g^2/\sigma^2 N \rangle$, the root-mean-squared pair separation fluctuation, $\langle \Delta_{B,} \rangle$, the reduced specific heat per bead, $\langle C_v^*/N \rangle$, and the reduced internal energy per bead, $\langle E^*/N \rangle$, as a function of the reduced temperature, $T^*$, for the $\beta$–double-hairpin at four selected values of the bias gap, (a) $g = 0.3$, (b) $g = 0.7$, (c) $g = 0.9$, and (d) $g = 1.3$.

**FIGURE 3** Phase diagram for the $\beta$–double-hairpin model as a function of the reduced temperature, $T^*$, and the bias gap, $g$.

**FIGURE 4** Final chain conformations of the $\beta$–double-hairpin (a) for $g = 0.7$ at $T^* = 0.3$ (the partially-ordered globule state), (b) for $g = 0.7$ at $T^* = 0.6$ (the ordered globule state), (c) for $g = 1.3$ at $T^* = 0.5$ (the ordered globule state), and (d) for $g = 1.3$ at $T^* = 0.2$ (the native state).

**FIGURE 5** The average value of the fraction of global energy minimum contacts, $\langle Q \rangle$, for the $\beta$–double-hairpin as a function of the reduced temperature, $T^*$, and the bias gap, $g$.

**FIGURE 6** (a) The Lindemann disorder parameter, $\langle \Delta_L \rangle$, for the $\beta$–double-hairpin as a function of the reduced temperature, $T^*$, for two selected values of the bias gap, $g = 0.3$ and $1.3$. (b) The Lindemann disorder parameter for an individual bead, $\Delta_{Li}$, for the $\beta$–double-hairpin as a function of the reduced distance of bead $i$ from the origin, $|\mathbf{r}_i|/\sigma$, for $g = 0.3$ at $T^* = 0.1$, $0.3$, and $0.6$. (c) Same as (b) but for $g = 1.3$ at $T^* = 0.2$, $0.6$, and $0.8$.

**FIGURE 7** The energy distribution in the transition region for the $\beta$–double-hairpin as a function of energy (a) for $g = 0.9$ at $T^* = 0.7949$ (the disordered-to-ordered globule transition), (b) for $g = 0.9$ at $T^* = 0.3998$ (the folding transition), and (c) for $g = 0.3$ at $T^* = 0.5848$ (the disordered-to-partially-ordered globule transition).

**FIGURE 8** The average values of the reduced squared radius of gyration per bead, $\langle R_g^2/\sigma^2 N \rangle$, the root-mean-squared pair separation fluctuation, $\langle \Delta_{B,} \rangle$, the reduced specific heat per bead, $\langle C_v^*/N \rangle$, and the reduced internal energy of system per bead, $\langle E^*/N \rangle$, as a function of the reduced temperature, $T^*$, for the $\beta$–clip model at (a) $g = 0.3$ and (b) $g = 1.3$, and for the $\beta$–twist model at (c) $g = 0.3$ and (d) $g = 1.3$.

**FIGURE 9** The Lindemann disorder parameter, $\langle \Delta_L \rangle$, as a function of the reduced temperature, $T^*$, and the bias gap, $g$, for (a) the $\beta$–clip and (b) the $\beta$–twist.

**FIGURE 10** The Lindemann disorder parameter for an individual bead, $\Delta_{Li}$, as a function the reduced distance of bead $i$ from the origin, $|\mathbf{r}_i|/\sigma$, for (a) the $\beta$–clip with $g = 0.9$ at $T^* = 0.1$, $0.26$, $0.5$, and (b) the $\beta$–twist with $g = 0.7$ at $T^* = 0.1$, $0.28$, $0.6$.

**FIGURE 11** Phase diagram as a function of the reduced temperature, $T^*$, and the bias gap, $g$, for (a) the $\beta$–clip model and (b) the $\beta$–twist model.

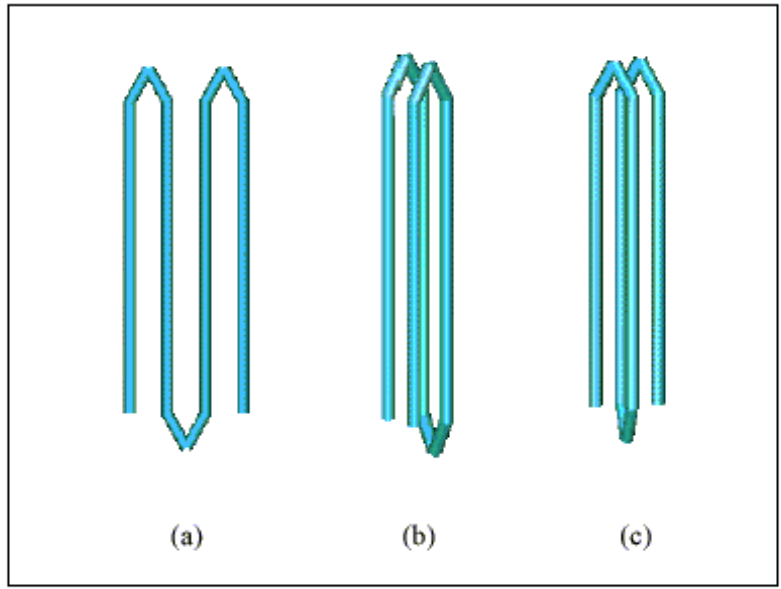

(a)        (b)        (c)

Fig. 2

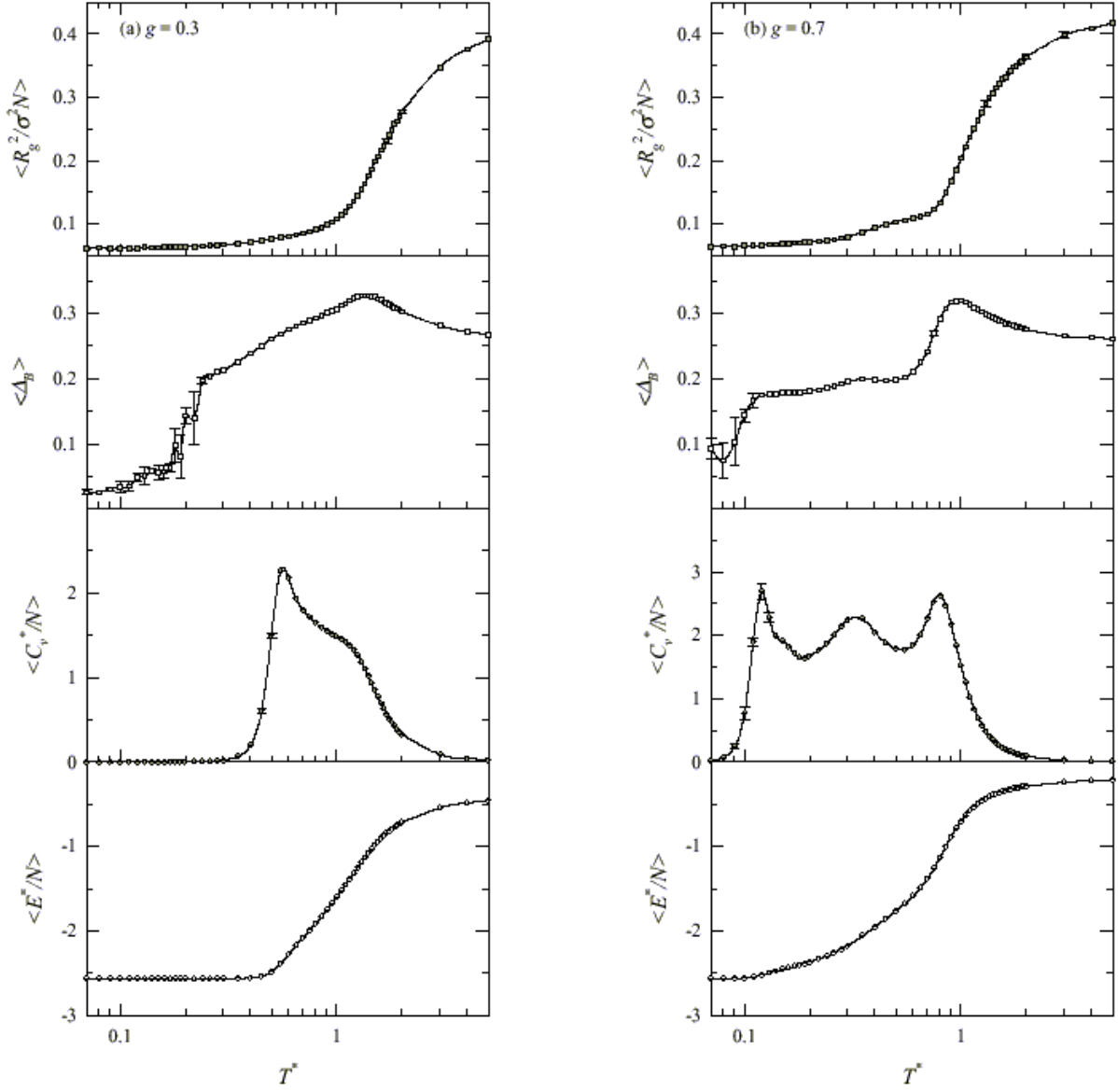

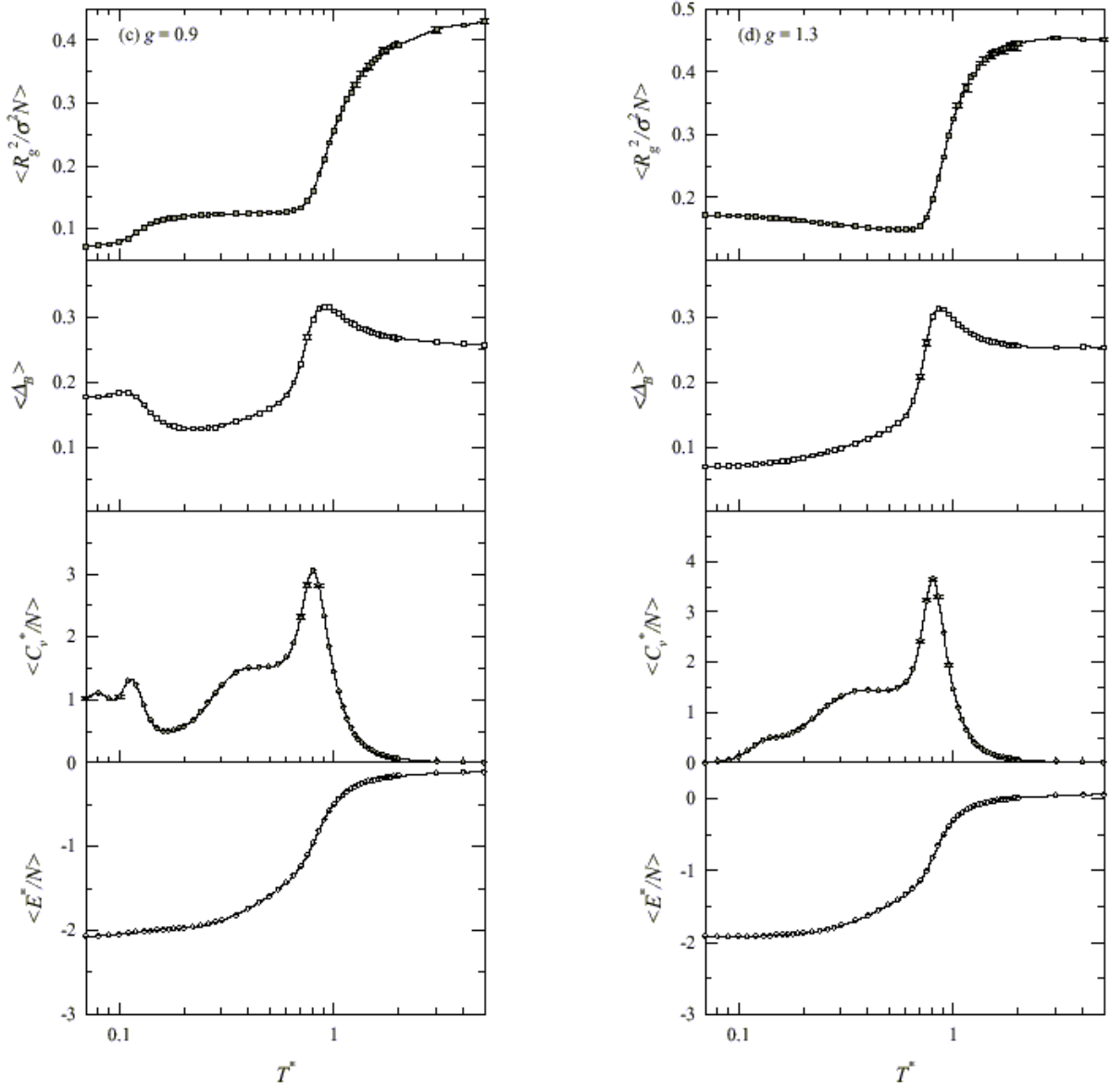

Fig. 3

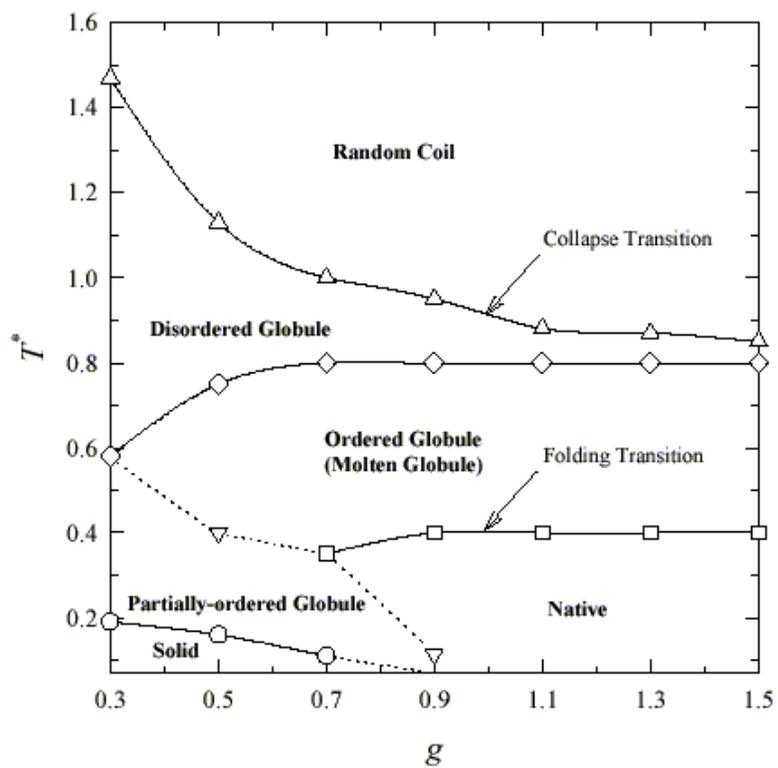

**Fig. 4**

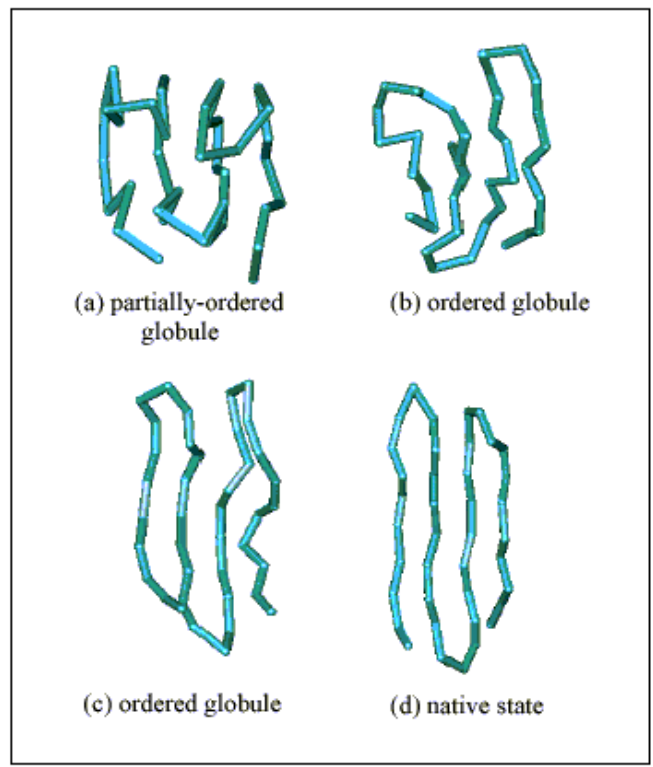

(a) partially-ordered globule

(b) ordered globule

(c) ordered globule

(d) native state

Fig. 5

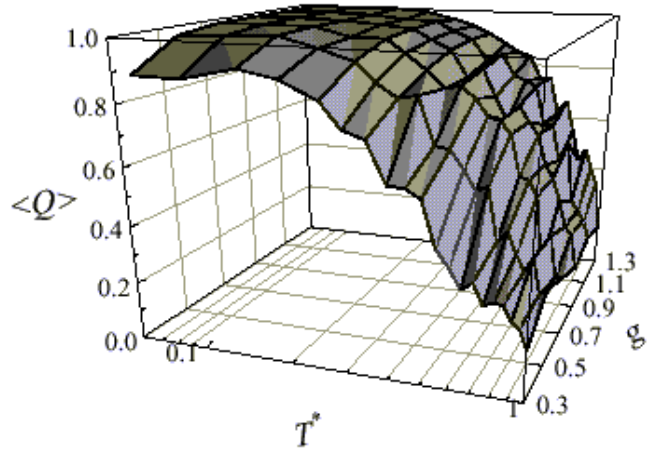

Fig. 6

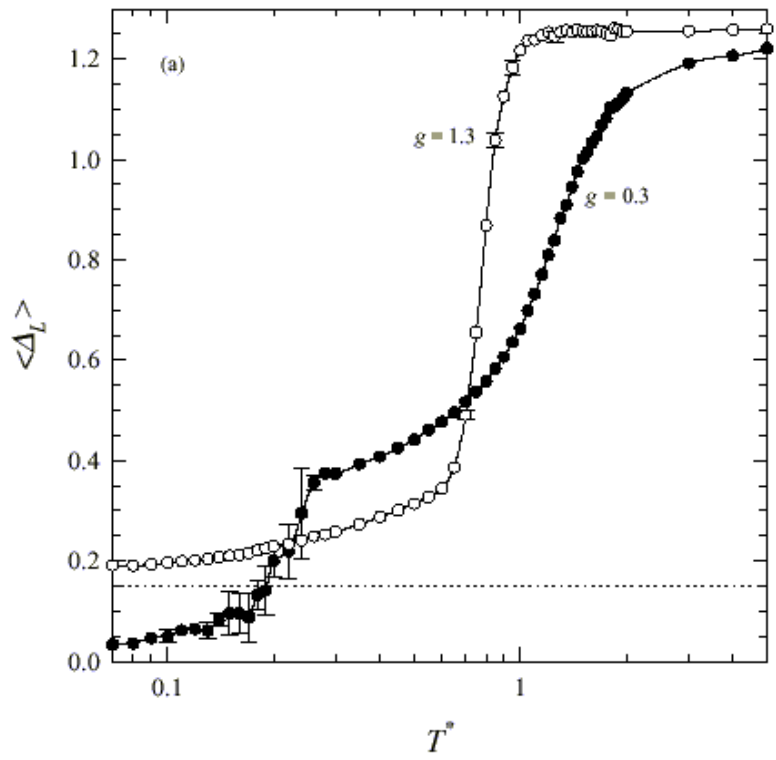

Fig. 6

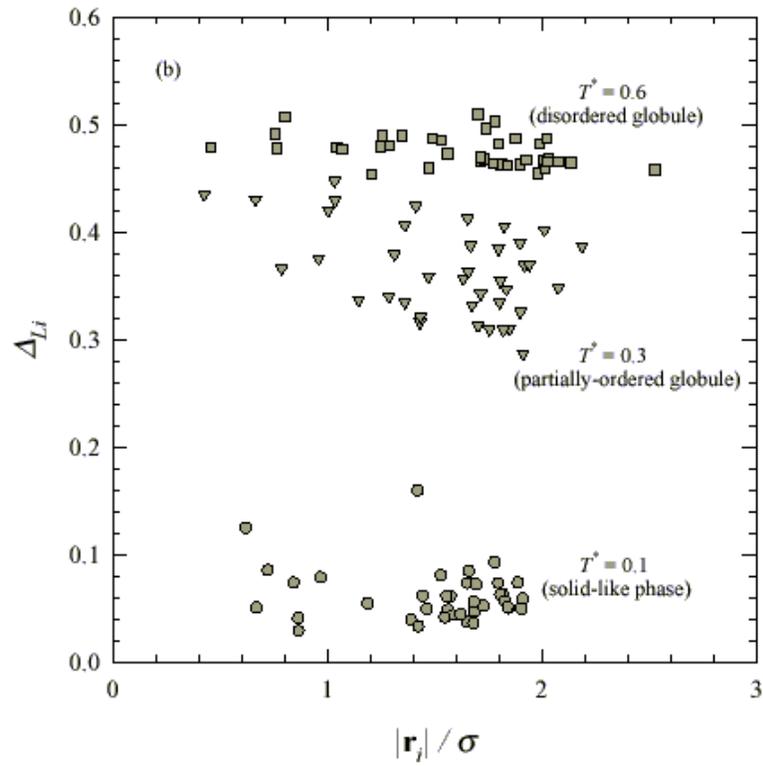

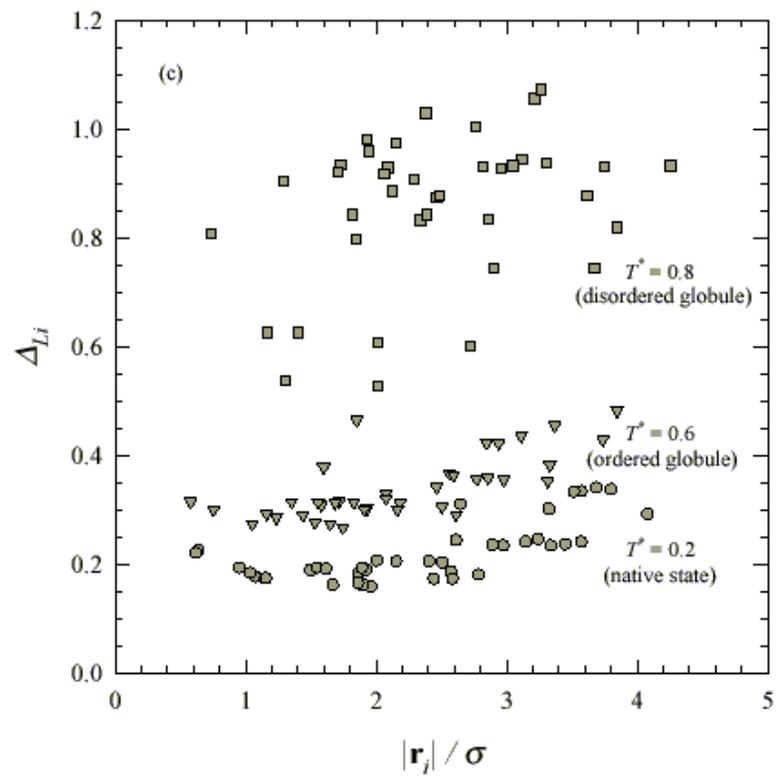

Fig. 7

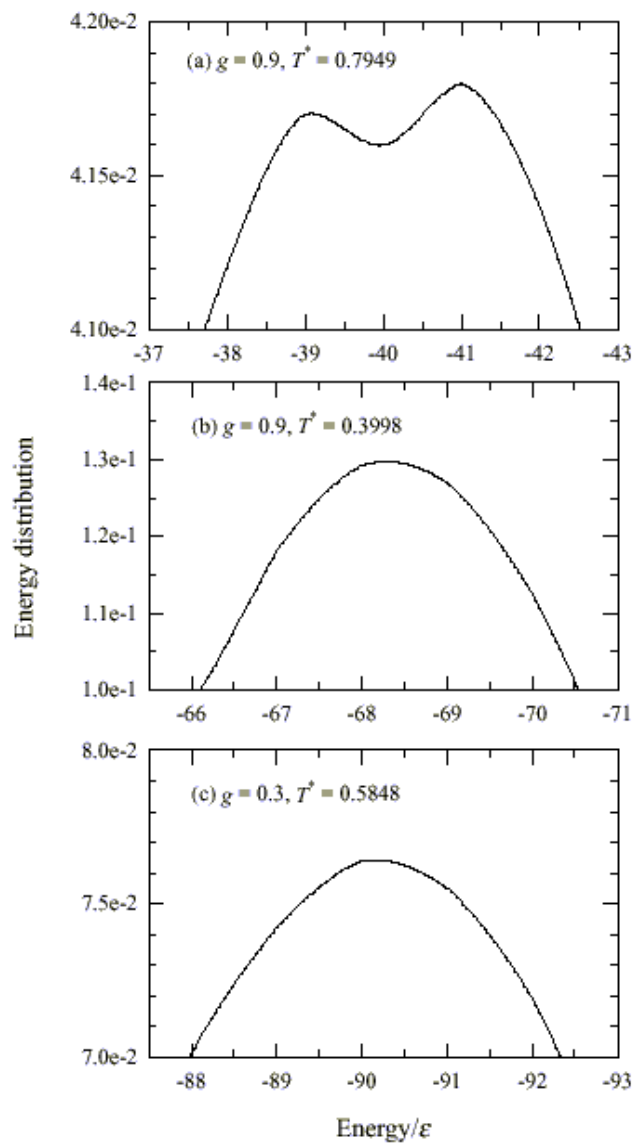

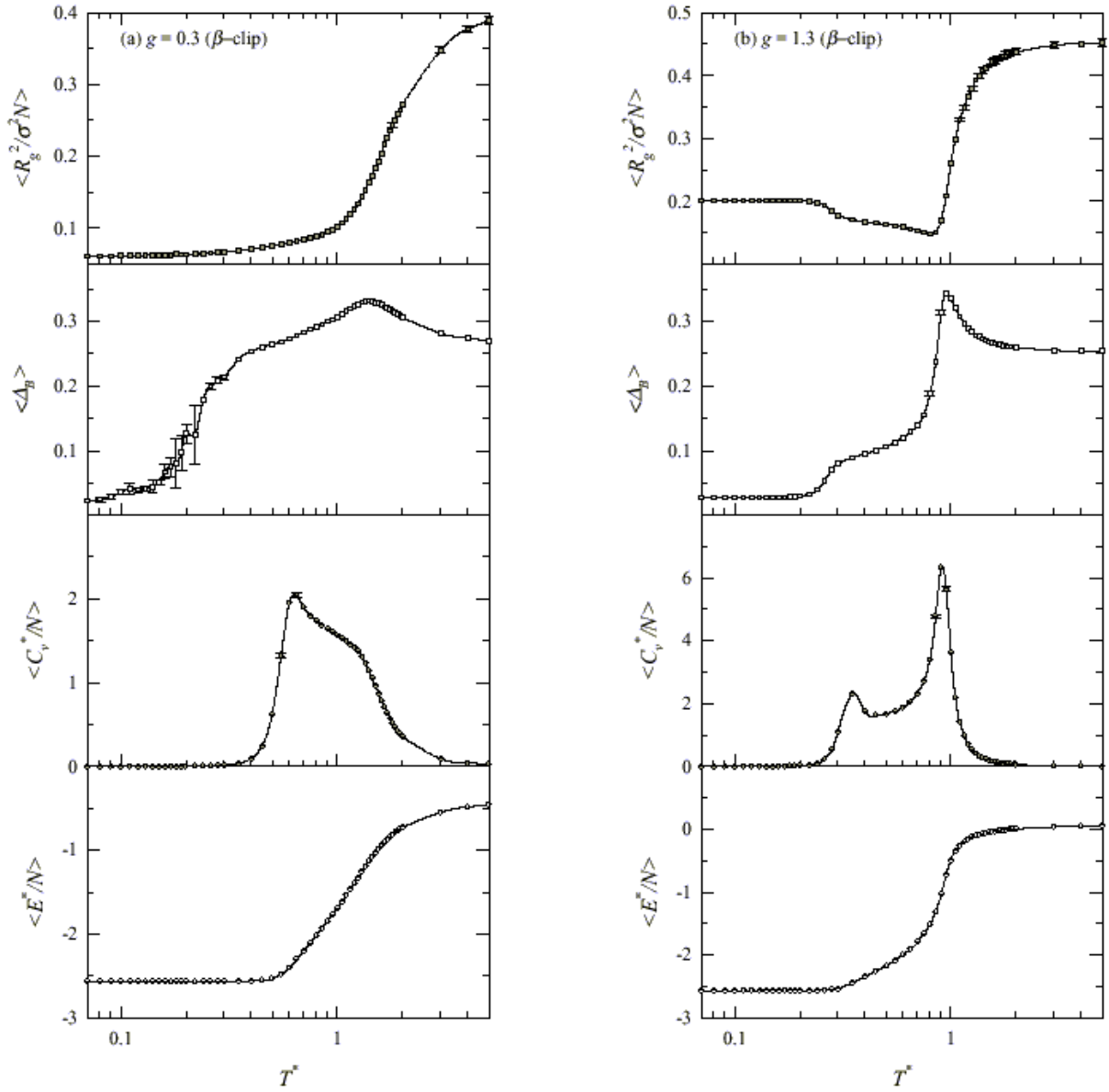

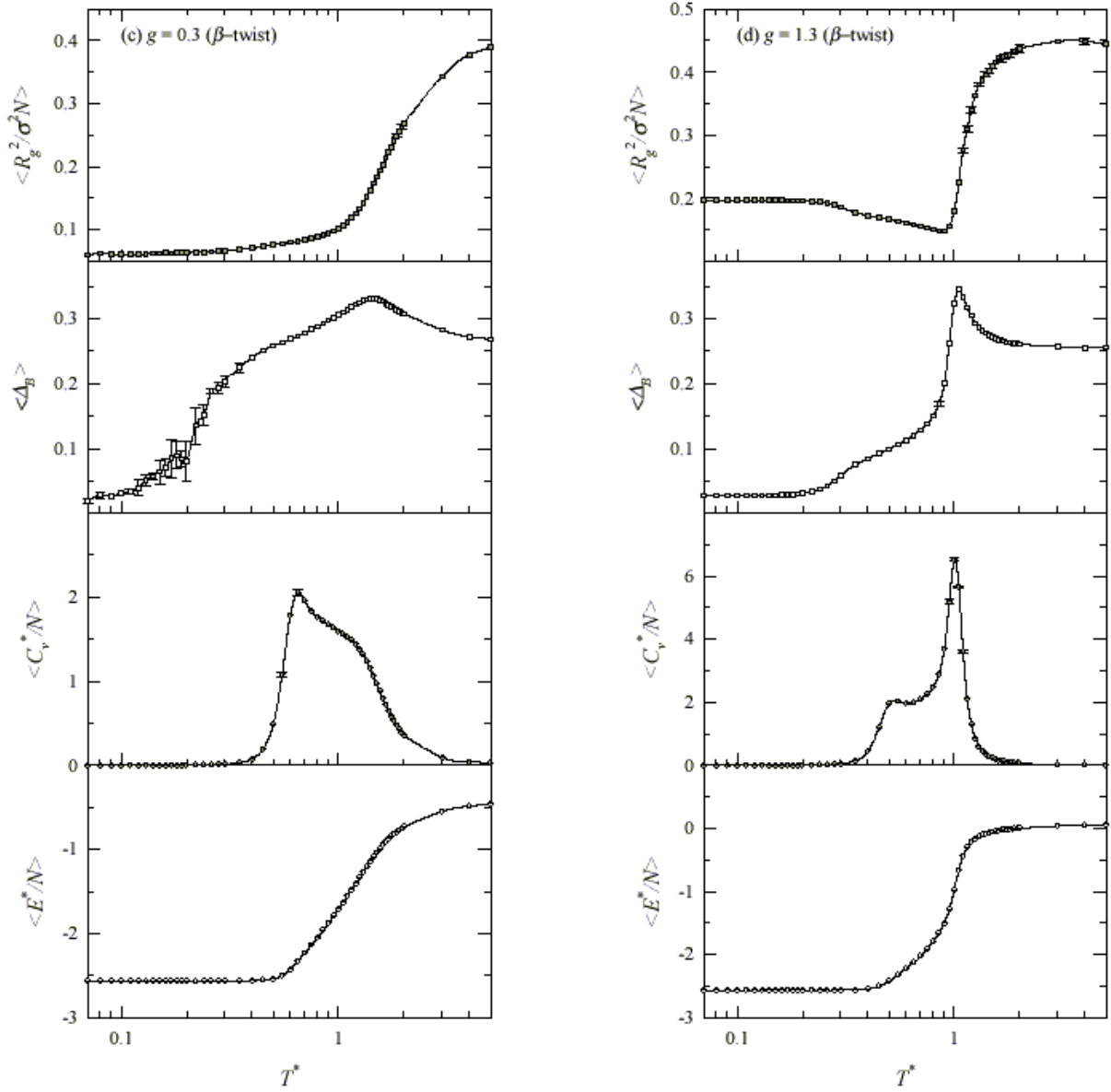

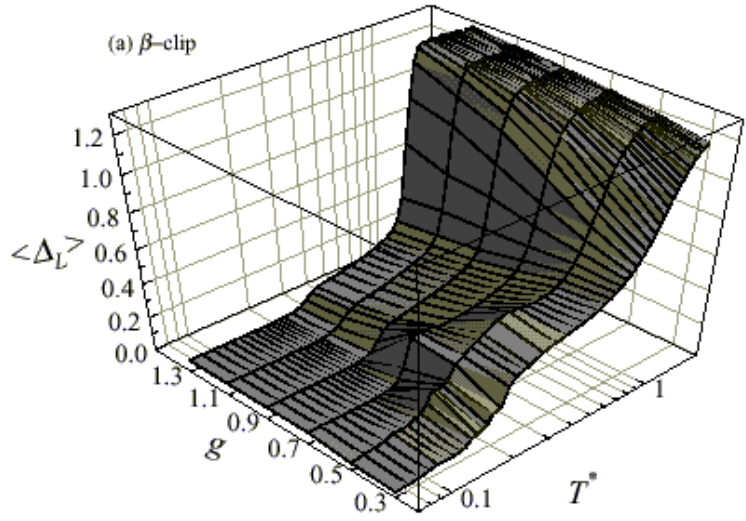

(a) β–clip

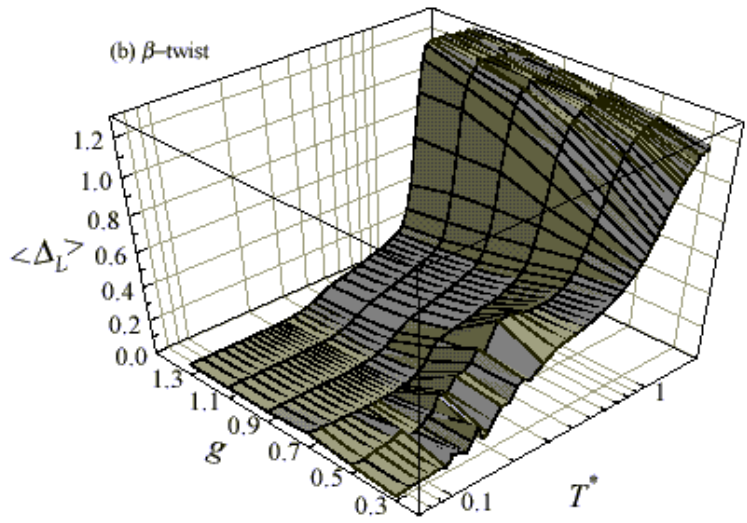

(b) β–twist

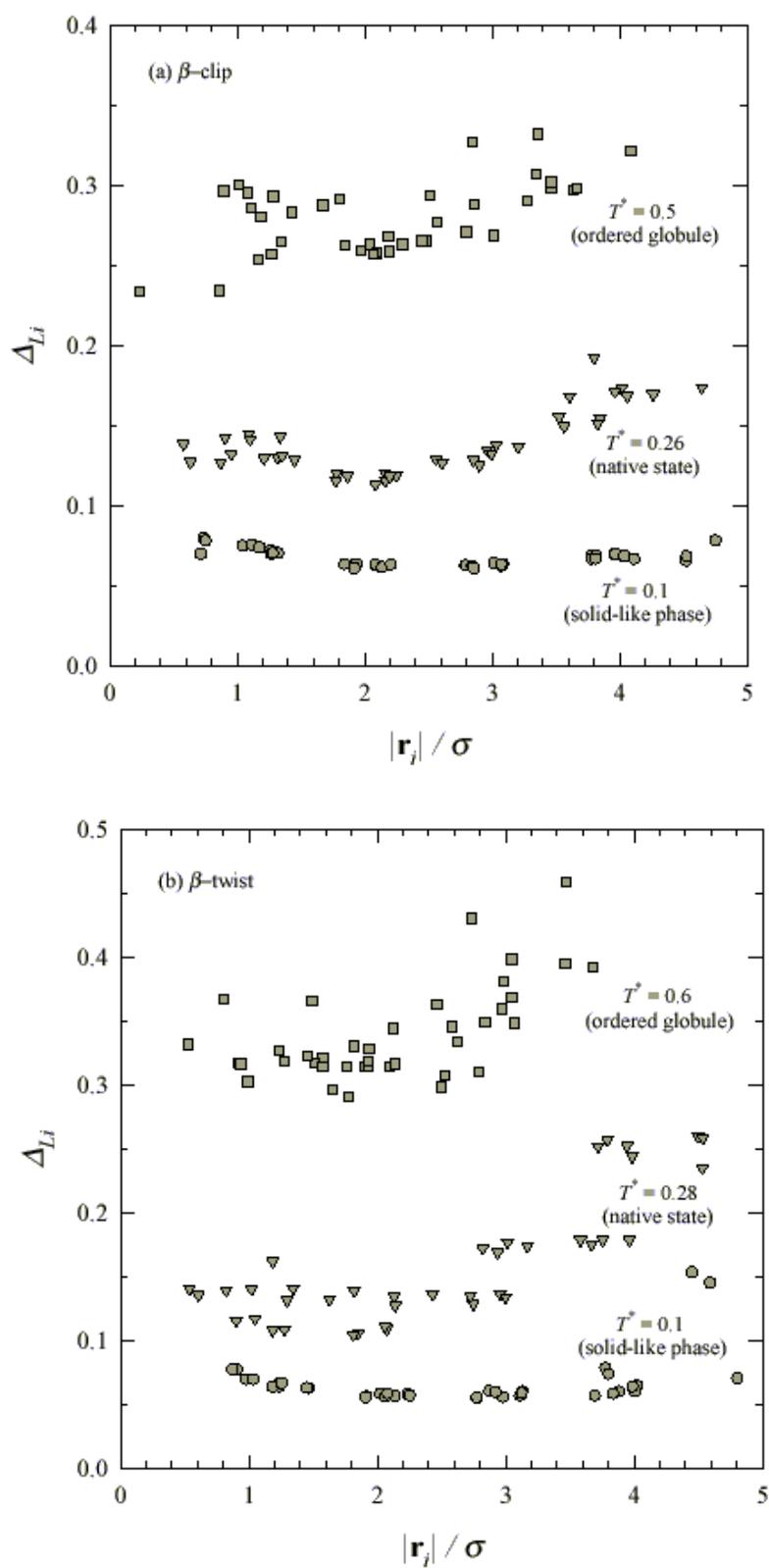

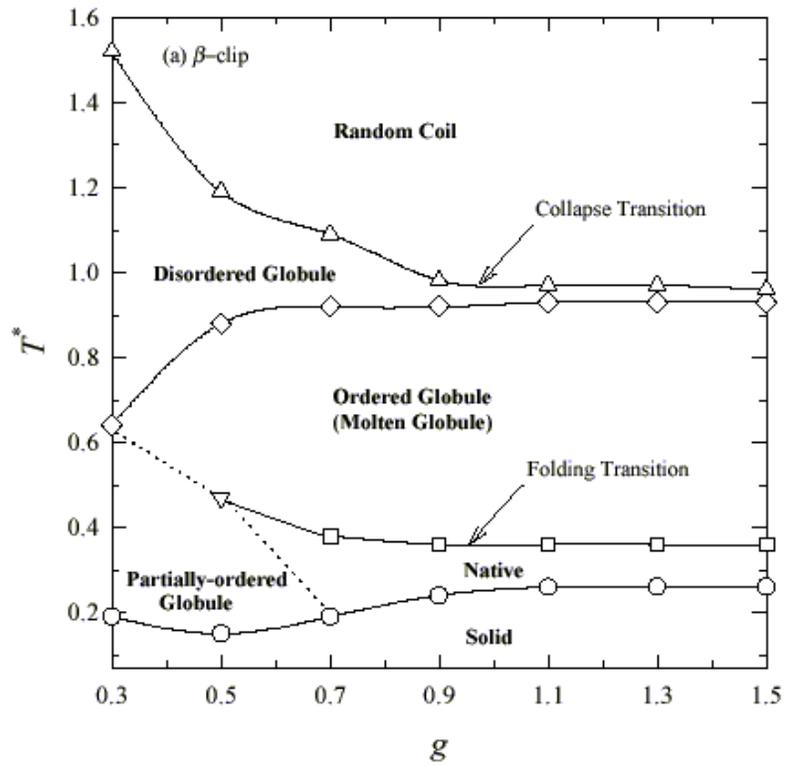

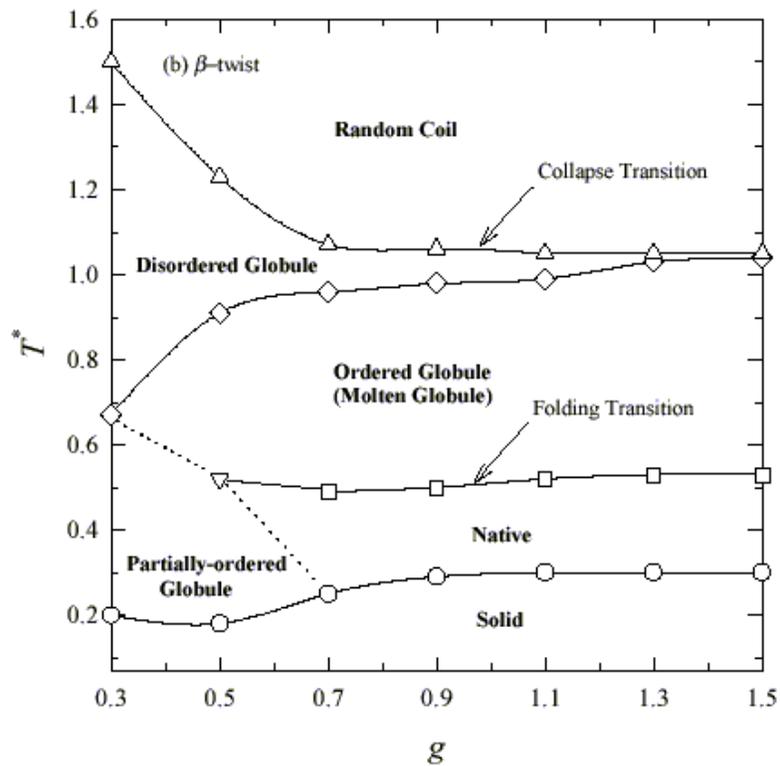